\documentclass[prl,amsmath,twocolumn, showpacs, superscriptaddress,10pt]{revtex4-1}
\usepackage{graphicx,color}
\usepackage{subfigure}
\usepackage{amsmath}
\usepackage{amssymb}
\usepackage{latexsym}
\usepackage{multirow}
\usepackage{bm}% bold math

\newcommand{\eeta}{{\boldsymbol{\eta}}}

\def\bnabla{\mbox{\boldmath $\nabla$}}

\begin{document}

\title{Non-equilibrium tuning  of the thermal Casimir effect}
\author{David S. \surname{Dean}} \affiliation{Univ. Bordeaux and CNRS,
  Laboratoire Ondes et Mati\`ere d'Aquitaine (LOMA), UMR 5798, F-33400
  Talence, France} \author{Bing-Sui \surname{Lu}}
\affiliation{Department of Theoretical Physics, J. Stefan Institute
  and Department of Physics, Faculty of Mathematics and Physics,
  University of Ljubljana, SI-1000 Ljubljana, Slovenia}
\author{A. C. \surname{Maggs}} \affiliation{UMR Gulliver 7083 CNRS,
  ESPCI ParisTech, PSL Research University, 10 rue Vauquelin, 75005
  Paris, France} \author{Rudolf \surname{Podgornik}}
\affiliation{Department of Theoretical Physics, J. Stefan Institute
  and Department of Physics, Faculty of Mathematics and Physics,
  University of Ljubljana, SI-1000 Ljubljana, Slovenia}

\begin{abstract}

In net-neutral systems correlations between charge
  fluctuations generate strong attractive thermal Casimir forces and
  engineering these forces to optimize nano-device performance is an
  important challenge. We show how the normal and lateral
  thermal Casimir forces between two plates containing Brownian
  charges can be modulated by decorrelating the
  system through the application of an electric field, which
  generates a non-equilibrium steady-state with a constant current in
  one or both plates,   reducing the ensuing fluctuation-generated
  normal force while at the same time generating a lateral drag
  force. This hypothesis is confirmed by detailed numerical
  simulations as well as an analytical approach based on
  stochastic density functional theory.
\end{abstract}


\maketitle
Electromagnetic (EM) fluctuation-induced interactions are dominant in
micro-electro-mechanical systems (MEMS)~\cite{bal07}, and their
presence is often viewed as undesirable as they engender stiction
between the micron-sized components of the MEMS devices. Controlling, or engineering, these forces is,
however, difficult as although they are all of electromagnetic origin,
they have contributions from both thermal and quantum fluctuations as
well as from different microscopic charge multipoles, e.g.\ ubiquitous
dipoles~\cite{mos97} and sometimes also free
monopoles~\cite{Drosdoff}.

Lifshitz~\cite{lif56} reformulated and generalized the original
zero-point electromagnetic field theory of idealized conducting
plates, proposed by Casimir~\cite{cas48,mos97,mil01,dal11}, in terms
of the dielectric and magnetic
permeabilities 
of real
materials 
sampled at all Matsubara
frequencies 
including a thermal zero frequency contribution. The Lifshitz formula for
EM field fluctuation-induced forces in symmetric interaction
configurations between standard materials 
reveals this interaction is generically attractive~\cite{par06}. Since
the interaction depends on frequency dependent material response
properties, it also suggests an immediate means of modulating or even
designing the Casimir force by appropriate changes in the material's
properties.  While this line of reasoning has been followed
successfully in meta-materials, it may be more useful to have a means
of switching EM fluctuation induced interactions directly {\em
  in-situ}.
A switch-like induced change in the optical properties of a material
indeed yields experimentally measurable differences in the interaction
between bodies in a number of cases: e.g.\ light or laser sources can
modify the charge carrier densities in semi-conductors
\cite{arn79,che07}, or induce phase
changes~\cite{tor10}. 
Theoretically it has been shown that the quantum Hall effect modified
conductivity can also be used to modify Casimir forces between
graphene sheets \cite{tse12}. Holding interacting materials at different temperatures
also allows modficatations of Casimir interactions~\cite{dor98,ant05,ant08,bim09,kru11a,kru11b,mes11,rod11,lu15}.

An alternative to the Lifshitz field-based formulation is presented by
the Schwinger matter-based approach where the Casimir force originates
in interactions between fluctuating charges and currents
\cite{mil01,jaf05,cug12}. Within this conceptual framework the
attraction between materials can be understood as being due to
correlations between microscopic EM source charge fluctuations that in
general reduce their (free) energy.
This implies that the effect of non-equilibrium driving the sources
with an external electric/magnetic field will {\em scramble\/} the
system's ability to develop charge correlations and could thus in
general reduce the attraction between the materials. The scenario of
engineering the EM fluctuation interactions by applying external
driving fields to MEMS is relatively easily implemented, compared to
the switching mechanisms based on material properties modifications
discussed above, and thus may be a promising technological direction
worth pursuing in detail.

In order to test the non-equilibrium driving hypothesis and assess its
ramifications, we analyze the system of two parallel conducting
plates, where only one of them is subjected to a current-inducing
applied external
electric field in a closed circuit configuration. For systems with
currents the usual methods of equilibrium statistical mechanics do not
apply.  We propose a methodology to study the effect of an imposed
current in the non-equilibrium steady state configuration on both the normal an lateral forces. The response of these forces to driving is surprisingly rich and this study thus opens up new perspectives for direct {\sl in situ\/} control of the EM fluctuation interactions.

\begin{figure}[h!]
\begin{center}
  \includegraphics[width=8cm,clip]{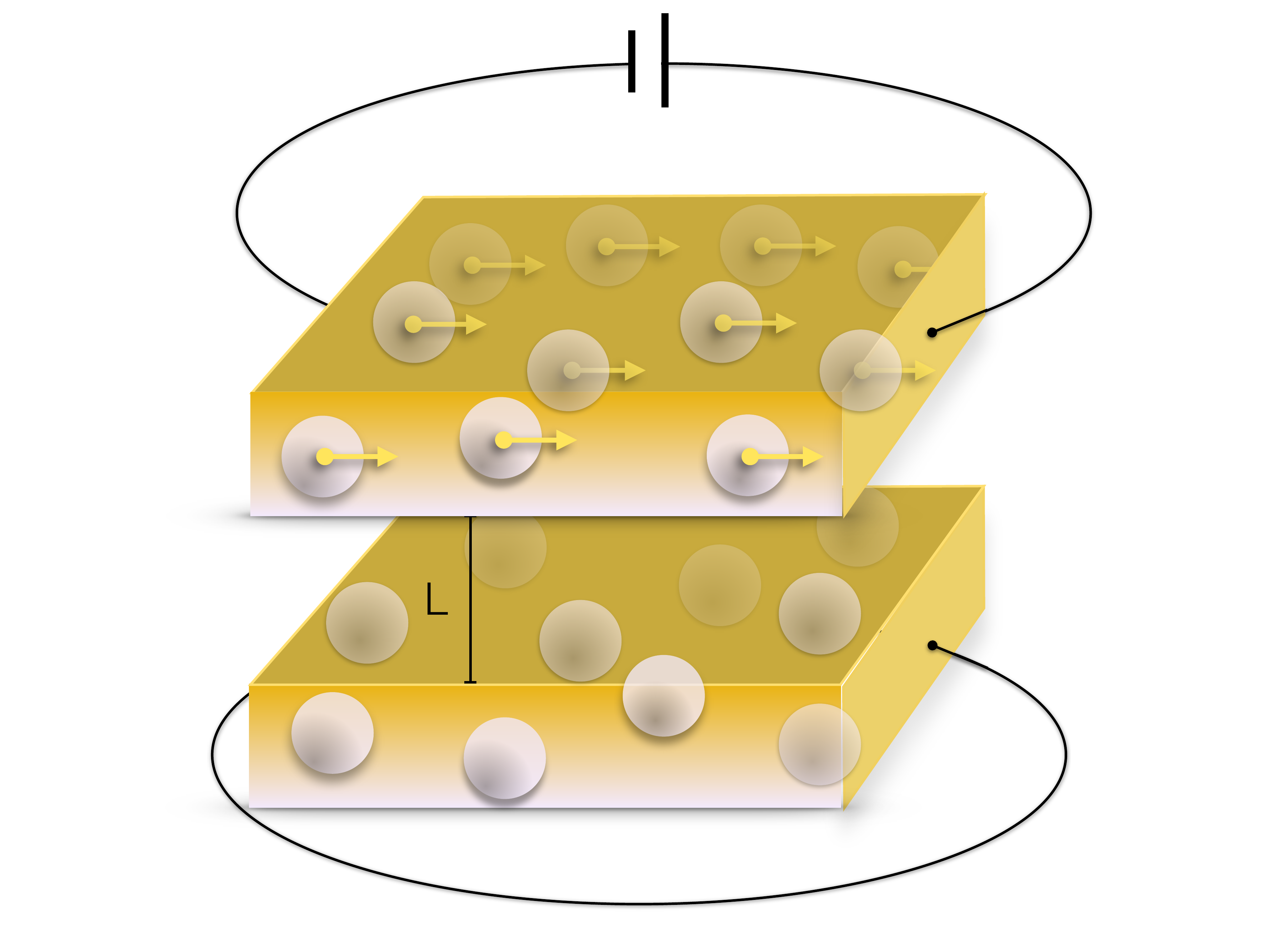}
  \caption{(color online) Schematic depiction of plates (1 and 2) containing mobile charges, with an external electric field ${\bf E}_{01}$ applied to plate 1 in a closed circuit configuration, driving a current flow in the plate set up using a battery. The circuit {\em of both plates} \cite{comm1} is open so charge flows through the system rather than accumulating at the edges of the plates.}\label{schematic}
\end{center}
\end{figure}

We analyze a well defined classical 2D jellium model, which can be
studied out-off equilibrium both numerically and analytically,
composed of two parallel plates with mobile charged Brownian particles
embedded within a uniform background charge
sheet~\cite{lev99,bue05,jan05}, see Fig.~(\ref{schematic}). In
equilibrium, at high temperatures in the weak coupling limit, the two
plates exhibit the universal thermal Casimir force
$F_\perp = -\frac{TS \zeta(3)}  {8\pi L^3}$ at large inter-plate
separations $L$,
with $S$ denoting the area of the plates, $\zeta$ is the Riemann zeta
function, and $T$ is the temperature of the two plates, assumed to be
the same for both. To explore the effect of an electric field on one
of the plates we need to specify the dynamics of the charges and we
adopt a Langevin model \cite{dea14,lu15} where a charge in the plate $\alpha$ at the
point $\bf X$ obeys the over-damped Langevin equation
\begin{equation} \frac{d {\bf X} } {dt} = \beta D_\alpha q_{\alpha}{\bf
    E}_{||\alpha}({\bf X}) + \sqrt{2D_\alpha}\eeta_{\alpha},
\label{lang}
\end{equation}
where $\eeta_{\alpha}$ is a zero-mean Gaussian white noise with
correlation function
$\langle \eta_{\alpha i}(t) \eta_{\alpha' j}(t')
\rangle=\delta_{\alpha\alpha'}\delta_{ij}\delta(t-t')$
and $\bf E_{||\alpha}({\bf X})$ is the local in plane electric field
in the plate $\alpha$, generated by the electric charge distributions
in both plates as well as any externally applied electric field. In
addition, $D_\alpha$ denotes the diffusion constant of the charges and
$\beta$~\cite{comm2} the inverse temperature so that $\beta D_\alpha$
is the local mobility, as deduced from the Einstein relation. Finally,
$q_\alpha$ is the charge of the mobile Brownian particles in the plate
$\alpha$ and if $\overline n_\alpha$ denotes the average density of
charge carriers in plate $\alpha$; the the uniform neutralising
surface charge density is thus
$\sigma_\alpha =- \overline n_\alpha q_\alpha$. The idealized model
above is amenable to both detailed numerical as well as analytical
developments, confirming our basic hypothesis that the driving field
modifies the charge correlation between the plates, thus leading to a
modified thermal Casimir force in direction normal as well as lateral
to the plates.

\noindent{\it Numerical simulations:\/} The two-plate system was
simulated by integrating the Langevin equations Eq. (\ref{lang}) for Brownian
particles \cite{SM}; the electrostatic forces due to the charges are computed
via Ewald summation. We use a non-cubic box of dimensions
$H\times H\times 3H$.\ with periodic boundary conditions in each
direction. We studied several separations between the plates up to a
maximum distance of $L=0.12H$; beyond this distance the interaction
between plates can be shown to cross-over to an exponential form due
to the discrete Fourier modes within the simulation box. In this
non-cubic geometry the undesired interactions between images of plates
in the z-direction are known to be negligible~\cite{yeh}. We worked
with a variable total number of particles, $N=1000$, $N=2000$, $ N=4000$ in
order to control finite size errors in the simulation. Apart from the applied
electric field ${\bf E}_{01}$, causing a current to flow within 
plate 1, the two plates are identical (the charges are identical and of the same number $N/2$ in each plate, the diffusion constants and temperatures are also the same). Plate separations are taken such that we find the far field universal Casimir effect at zero applied field.

The normal and lateral force acting on the plate $\alpha$ are computed
from the formulas
\begin{equation}
  {\bf F}_{\perp,|| ~\alpha}(L) =  q_\alpha\int_{S_\alpha} d^2{\bf x}\
  {\bf E}_{\perp,|| ~\alpha}({\bf x}) 
  ~\Delta n_\alpha({\bf x}),\label{force}
\end{equation}
where
$n_\alpha({\bf x})= \sum_{i}\delta({\bf X}_i-x)$ is the density of mobile charge carriers and
$\Delta n_\alpha({\bf x}) = n_\alpha({\bf x})-\overline n_\alpha$ is  the fluctuation
about its spatially averaged value $\overline n_\alpha$, which is the same as that of  
the neutralizing uniform background charge. In Eq. (\ref{force})  $\perp, ||$ indicates
the direction with respect to the bounding plates.  The electrostatic
potential $\phi({\bf x}) $ in the plate
$\alpha$ 
has contributions both from the plate $\alpha$ as well as plate
$\alpha'$ (opposite) mediated
by the standard Coulomb interaction,
while the dielectric constant $\epsilon$ is assumed to be homogeneous.

\begin{figure*}[t]
\begin{center}
\begin{minipage}[b]{0.45\textwidth}
\includegraphics[width=\textwidth]{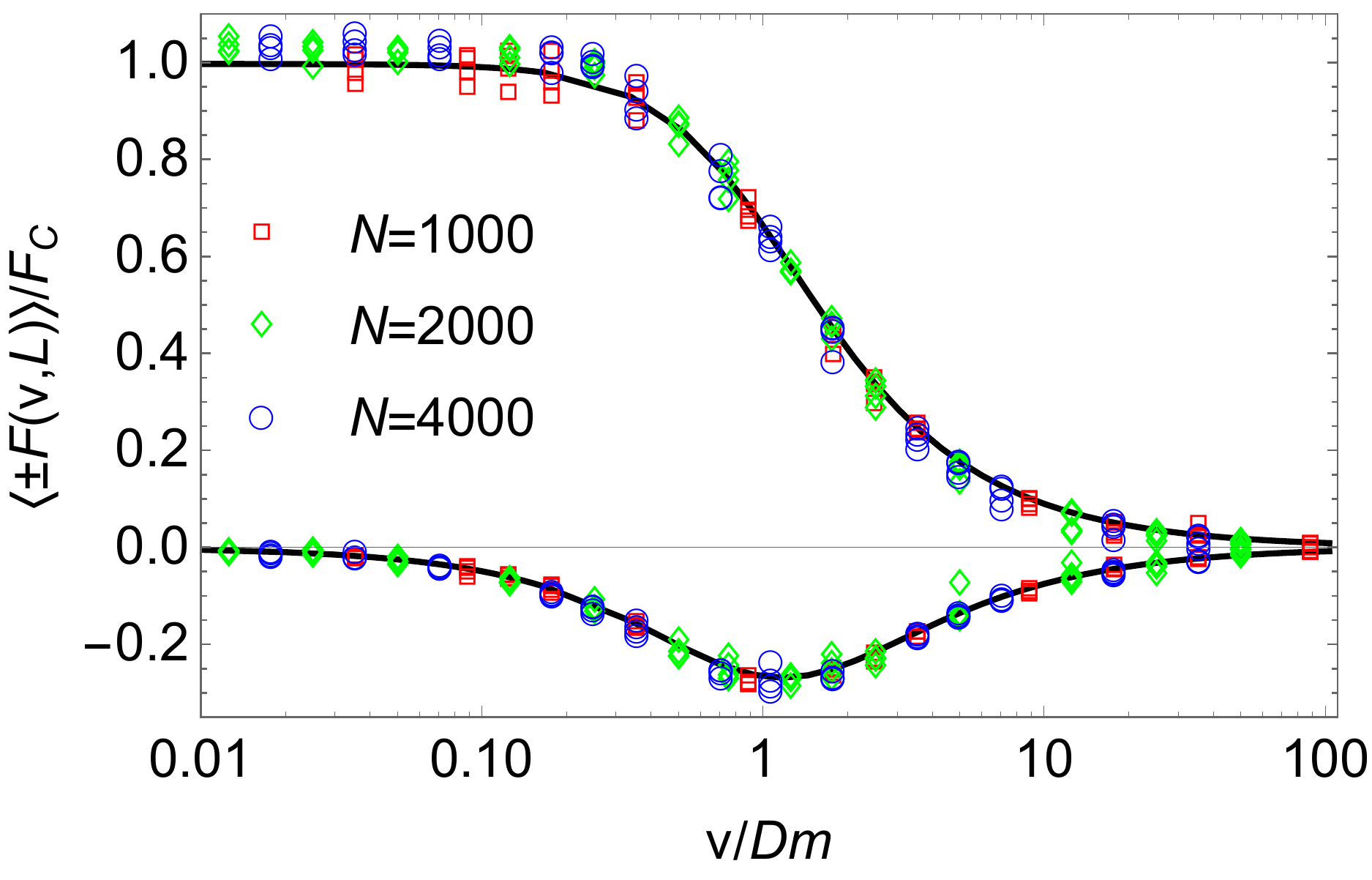} (a)
\end{minipage} 
\begin{minipage}[b]{0.45\textwidth}
\includegraphics[width=\textwidth]{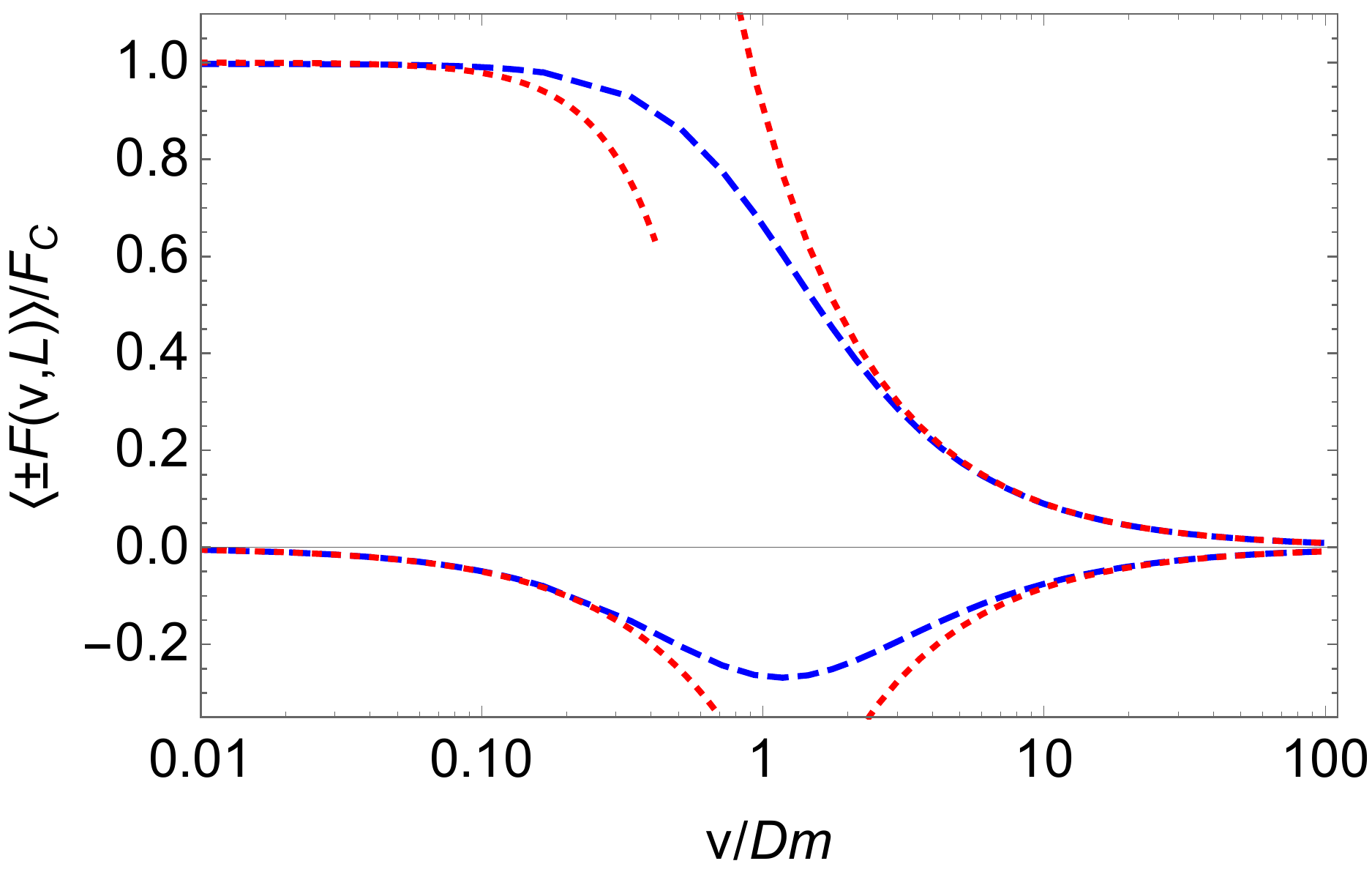} (b)
\end{minipage}
\end{center}

\caption{(color online) (a): Evolution of the amplitude of the perpendicular and transverse components of the Casimir force with the driving field, normalized by the thermal Casimir force, $F_{C}=-TS\zeta (3)/8 \pi L^3$. For ease of comparison,  we have displayed the component of the normalized tangential force that is parallel to the driving field, with a sign opposite to that of the normalized perpendicular force. Simulation results are taken from systems with $N=1000$ ({red squares}), $N=2000$ ({green diamonds} ) and $N=4000$ ({blue circles}).  For each value of $N$ the interaction is evaluated for four separations: $L = 0.02H$, $L = 0.05H$, $L = 0.08H$, $L = 0.12H$.  The vertical data spread correspond to residual systematic, finite size errors in the simulations.  The theoretical predictions, Eqs.~(\ref{avf1},~\ref{avf2}), for $L \times m=2000$ are shown by black solid curves. % 
(b): Evolution of the amplitude of the perpendicular and transverse components of the Casimir force between identical plates ($D_1=D_2=D$, $m_1=m_2=m$) with the driving field, normalized by the universal thermal Casimir force.Theoretical predictions, Eqs.~(\ref{avf1},~\ref{avf2}), for $Lm=2000$ are shown by blue dashed curves, whilst the corresponding small field and large field asymptotes, Eqs.~(\ref{eq:fperp},~\ref{eq:fperplarge},~\ref{eq:aspar},~\ref{eq:ftangsmall}), are shown as the red, dotted curves.}
 \label{fig:force}
 \end{figure*}

The numerical results for the average of the two forces are plotted
in Fig.~\ref{fig:force}.  They are both fluctuational in nature, and in principle the statistics of the force can be measured. For small driving fields the
normal Casimir force saturates to the equilibrium thermal Casimir
force, while the lateral force vanishes in equilibrium. As the field
increases there is a {\sl monotonic\/}  decrease in the
amplitude of the normal force, that eventually asymptotes to zero, and
a {\sl non-monotonic\/} variation of the lateral force that is zero
for small as well as large values of the driving field.  

\noindent{\it Dynamical density functional theory\/} The density
fields $n_\alpha({\bf x},t)$ in each plate evolve according to the
exact stochastic partial differential equation~\cite{kaw94,dea96}
\begin{align}\frac{\partial n_\alpha({\bf x},t)}  {\partial t} &=
  D_\alpha\nabla_{||} \cdot\left[ \nabla_{||} n_\alpha({\bf x},t)
    -\beta_\alpha q_\alpha{\bf E}_{||\alpha} n_\alpha({\bf
      x},t)\right] \nonumber \\&+ \nabla_{||}\cdot[ \sqrt{2 D_\alpha
    n_{\alpha}({\bf x},t)} \eeta_\alpha({\bf x},t)].\label{edyn}
\end{align}  
In this density representation of the dynamics, the noise
$\eeta_\alpha({\bf x},t)$ is a spatio-temporal Gaussian white noise
vector field of mean zero and with correlation function
$\langle \eta_{\alpha i}({\bf x},t)\eta_{\alpha' j}({\bf x}',t)\rangle
= \delta_{\alpha\alpha'}\delta_{ij}\delta(t-t')\delta({\bf x}-{\bf
  x}')$.

To make analytical progress we expand the deterministic term in
Eq. (\ref{edyn}) to linear order in the density fluctuations
$n_\alpha$ and the noise term to zeroth order (since it is of zero
mean this is consistent with the first order expansion of the
deterministic terms). This approximation has already been used to
examine interactions between plates out of equilibrium in e.g.
evolution to the equilibrium force for initially out of equilibrium
plates~\cite{dea14}, as well as for the non-equilibrium force between
plates held at different temperatures~\cite{lu15}. This small density
expansion was recently shown to reproduce  Onsager's classical
results on the conductivity of strong electrolyte solutions
\cite{hem96} in a very straightforward and compact manner
\cite{dem15}. Within the small density fluctuation approximation the
two dimensional Fourier transform of the density fluctuations, defined
as
$\Delta\tilde n_\alpha({\bf Q},t) =\int_{S_\alpha} d^2{\bf x} \ \exp(-i{\bf
  Q}\cdot {\bf x})\Delta n_\alpha({\bf x},t)$,
has a steady state correlation function
$\langle \Delta\tilde n_\alpha({\bf Q})\Delta \tilde n_{\alpha'}({\bf Q}')\rangle =
{(2\pi)}^2 \delta({\bf Q}+{\bf Q}') C_{\alpha\alpha'}({\bf Q})$
which obeys
\begin{equation}
  M_{\alpha\gamma}({\bf Q}) C_{\gamma\alpha'}({\bf Q}) + C_{\alpha\gamma}({\bf Q}) M_{\gamma\alpha'}^{{\rm T}}(-{\bf Q}) =2\delta_{\alpha\alpha'} D_\alpha \overline n_\alpha Q^2 .
  \label{gbacefjk}
\end{equation}
The matrix $M(\bf Q)$ is given by
\begin{equation}
  M_{\alpha\gamma}({\bf Q})\!=\!Q^2\!\left(\! \tilde D_\alpha \delta_{\alpha\gamma} \!+\!\beta  q_\alpha q_\gamma \overline n_\gamma D_\alpha \tilde G({\bf Q}, z_{\alpha\gamma}) \right),
\end{equation}
where $ \tilde D_\alpha = D_\alpha\!\left( 1\!-\!i\beta q_\alpha \overline n_\alpha {{\bf \hat Q}\cdot\bf{E}_{0\alpha}}/{Q}\right)$, $z_{\alpha\gamma} = z_\alpha-z_\gamma$ and ${\bf \hat Q}$ is the unit wave-vector. The term $\tilde G({\bf Q}, z_{\alpha\gamma})$ denotes the in-plane Fourier transform of the Coulomb interaction $G({\bf x}, z)$ (without the charge factors) and is given by $\tilde G({\bf Q}, z) = \exp(-Q|z|)/2\epsilon Q$.
The components of the
fluctuation force in Eq.~(\ref{force}) can be expressed as
\begin{equation}
  \langle F_{\perp} \rangle = -q_1 q_2 \! \int_{S_1} \!\!\!\!d{\bf x}\!\!\int_{S_2} \!\!\!\!d{\bf y} \langle \Delta n_1({\bf x}) \Delta n_2({\bf y}) \rangle \frac{\partial G({\bf x}-{\bf y},L)}{\partial L}
  \label{frc1}
\end{equation}
and
\begin{equation}
  \langle {\bf F}_{||} \rangle = - q_1q_2\int_{S_1}\!\!\!\!d{\bf x}\!\!\int_{S_2}\!\!\!\!d{\bf y} ~\langle  \Delta n_1({\bf x}) \Delta n_2({\bf y}) \rangle\bnabla_{||} G({\bf x}-{\bf y},L) .
  \label{frc2}
\end{equation}
The Fourier transform of
$\langle \Delta n_\alpha({\bf x}) \Delta n_\beta({\bf y}) \rangle$ can then be
obtained from Eq.~(\ref{gbacefjk}),
which together with the definitions Eqs.~(\ref{frc1}),~(\ref{frc2}) yield
the average normal and lateral force. Defining
\begin{equation}
f(Q, L) = \log{\Big( 1+\Delta({\bf v}_1,{\bf v}_2)^2  - \frac{m_1 m_2  \exp^{-2QL}}{(m_1+2Q)(m_2+2Q)}\Big)} \label{fq}
\end{equation}
with ${\bf v}_\alpha = \beta q_\alpha D_\alpha {\bf E}_{0\alpha}$ the average velocity of the mobile charges in the plate $\alpha$, $m_\alpha = \beta \bar{n}_\alpha q_\alpha^2 /\epsilon$ is the inverse screening length for a 2D Coulomb gas of mobile charges in equilibrium and $\Delta({\bf v}_1,{\bf v}_2) = 2 {({\bf\hat Q}\!\cdot\!({\bf v}_1\!-\!{\bf v}_2))}/{(D_1m_1\!+\!D_2m_2\!+\!2(D_1\!+\!D_2)Q)}$, we can write \cite{SM}

\begin{equation}
 \langle F_\perp(L)\rangle = -{\textstyle\frac12}T S \int \!\! \frac{d^2{\bf Q}} {{(2\pi)}^2}  \frac{\partial f(Q, L) }{\partial L},
 \label{avf1}
\end{equation}  
and
\begin{equation}
  \langle {\bf F}_{||}(L) \rangle = T S \int \!\! \frac{d^2{\bf Q}} {{(2\pi)}^2}  \frac{\partial f(Q, L) }{\partial L} {\bf \hat Q} ~\Delta({\bf v}_1,{\bf v}_2).
  \label{avf2}
\end{equation}

We note that when ${\bf v}_1-{\bf v}_2=0$ we recover the
expression for the equilibrium thermal Casimir force for
$\langle F_\perp(L)\rangle$, while the lateral force is zero, since in
the common rest frame of the moving charges the system is in
equilibrium. 

The normal force is monotonic in its variation with respect to the relative
difference of the bare velocity of the charges in each
plate. 
In the far field limit where $L \gg m_1^{-1},\ m_2^{-1}$
and when in addition $\vert {\bf v}_1-{\bf v}_2\vert \ll u_1,\ u_2$,
where $u_\alpha = D_\alpha m_\alpha$ define an intrinsic velocity
scale in each plate, the average force simplifies to give
\begin{equation}
  \langle F_\perp(L)\rangle \approx - \frac{T S}{8\pi L^3}\left[
    \zeta(3) - \frac{\pi^2 \vert {\bf v}_1-{\bf v}_2\vert ^2}{3{(u_1 +
        u_2)}^2} \right].
\label{eq:fperp}
\end{equation}
The effect of the applied field can thus be seen as renormalizing the
effective {\em Hamaker\/} constant associated with the $1/L^3$ power
law.  The far field fluctuation induced attraction between the plates
thus monotonically decreases upon increasing the relative velocity.
In the opposite limit of large relative
velocity, 
we find that the force decays as
\begin{equation}
  \langle F_\perp(L)\rangle \approx - \frac{TS}{32 \pi L^3}
  \frac{(u_1+u_2)}{| {\bf v}_1 - {\bf v}_2 |} \left [8 -\frac{\pi^2} {3} -8
    \ln{2} + 4\ln^2{2} \right ] 
    \label{eq:fperplarge}
\end{equation}
Contrary to the normal force, the lateral force is not monotonic in
the relative velocity and shows a well defined maximum.
On the two sides of this minimum the lateral force behaves as 
\begin{equation}
 \langle F_\parallel(L)\rangle \approx - \frac{TS}{16 \pi L^3}
 \frac{(u_1+u_2)}{|{\bf v}_1 -{\bf v}_2 |}  
\label{eq:aspar}
\end{equation}
in the large field limit,  and for small fields  as 
\begin{equation}
\langle F_\parallel(L)\rangle \approx -\frac{T S |{\bf{v}}_1-{\bf{v}}_2|}{8 \pi L^3 (u_1+u_2)} 
\left[ \zeta(3) - \frac{\pi^2 |{\bf{v}}_1-{\bf{v}}_2|^2}{128(u_1+u_2)^2} \right]. 
\label{eq:ftangsmall}
\end{equation}
A similar non-monotonic drag force has recently been predicted for single particles coupled to thermally fluctuating classical fields \cite{dem10,dem11}.

In Fig.~\ref{fig:force}a we compare the theoretical predictions, Eqs.~(\ref{avf1},~\ref{avf2}), for the normal and lateral forces with the results of our numerical simulations. We see that, despite the relatively low temperature of the system, the agreement for both forces is excellent.  The asymptotic results for the small and large field limits, Eqs.~(\ref{eq:fperp},~\ref{eq:fperplarge},~\ref{eq:aspar},~\ref{eq:ftangsmall}), are compared to the full numerical integration of Eqs.~(\ref{avf1},~\ref{avf2}) in  Fig.~\ref{fig:force}b. 

\noindent{\em Conclusions:\/} We have introduced a simple model
exhibiting the thermal Casimir effect and shown that when the system
is driven by an external electric field, the thermal Casimir force,  both its normal and lateral components, can be modulated in a controlled and reversible manner. The underlying physical mechanism is that the external driving electric fields suppress the charge correlations which are responsible for the fluctuation interaction. Indeed the Onsager study of the field dependence of electrolyte conductivity~\cite{hem96}, the Wien effect,
shows that the increase in conductivity is due to the fact that the
applied field suppresses Debye screening. The underlying mechanism
here is clearly related and we have clearly exhibited the effect in numerical simulations, and 
analytically, taking into account all the non-equilibrium physics in the model via its microscopic formulation. Indeed one of the prime advantages of the model studied here is that one can carry out numerical experiments to measure fluctuation induced interactions {\em out of equilibrium} and future study of this model should enable the study of fluctuation induced interactions beyond the weak density fluctuation approximation employed in our analytical study.

Extensions of this model to more realistic systems/models are clearly
desirable in order to test the general hypothesis on field induced
correlation scrambling put forward here. Another natural question to ask is whether non-equilibrium forcing can be used to amplify correlations and thus enhance the fluctuation induced attractive force? The Brownian conductor model
should be extended to take into account  inertial effects so
that it more closely resembles the Drude model. In addition,
retardation effects in the electromagnetic interactions between the
charges could also be incorporated. Ultimately one should consider
non-equilibrium quantum field theories in order to understand the
quantum aspects of the problem. Clearly AC driving fields also
constitute and interesting line of research, from both a numerical and
analytical point of view.

\begin{acknowledgments}{This work was partially financed by a Joliot Chair
  of the ESPCI (RP), the  ANR FISICS (DSD), the ANR FSCF (ACM), and ARRS P1-0055 (BSL).}
\end{acknowledgments}

\section{Supplementary Material}

\subsection{Numerical simulations}

We used the standard three-dimensional formulation of the Ewald
summation introducing a real-space range $\alpha$. The complementary
long-range part of the interaction is calculated in Fourier space,
requiring modes out to wave-vectors $q_m\sim 1/\alpha$. The 
interaction energy as well as the forces are independent of $\alpha$
which can then be used to optimize the speed of the
calculation. Balancing the effort in real and Fourier space leads to a
value of $\alpha$ which decreases with the number of simulated
particles so that the computer effort to calculate the force on all of
the particles scales as $O(N^{8/5})$, with $N$ the number of
particles. In practice, for each value of $N$, we made a series of test
runs with variable $\alpha$ chosen to produce the most rapid code.  We used
a relative precision of $10^{-11}$ in the Ewald evaluation.

We integrated the Langevin equation using the Euler method
with a time step $dt$. We chose this step as follows: First we
performed a series of short simulations with very large time steps to
find the stability limit of the integrator. We then divided this value
of $dt$ by 100 to find a stable regime where integration errors are
not too large. We  performed high statistics simulations for
different time steps about our first estimate for the production time
step. We measured the force between plates with statistical errors
better than 2\%, estimated using the method of \cite{errors}, and
found the value of $dt$ which also gives systematic errors of order of
2\%. In practice it is immediately clear when the step size has been
chosen too large when comparing several different simulations.

We note that the effect of a finite step size can be studied
analytically using the concept of {\it inverse error analysis\/}. A
set of variables $i$ with energy function $\mathcal{H}$ when simulated by a
Euler integrator generates an equilibrium measure characterised by an
effective energy \cite{reverse}
\begin{equation}
  \bar{\mathcal{H}} = \mathcal{H} + \frac{D dt}{4} \sum_i  \left (   2  \nabla_i^2 \mathcal{H} -
    \beta (\nabla_i \mathcal{H})^2 \right ),
\end{equation}
where $D$ is  the particle diffusion constant. When applied to a pair of
Coulomb particles this gives an extra effective potential in
$dt/r^4$. Remarkably this correction is very similar to the Wigner
expansion \cite{quantum} in quantum mechanics. The integration error
looks very much like the quantum Casimir interaction in the high
temperature limit, with $dt$ playing a role comparable to $\hbar^2$.

With the choice of the time step set as above, we  performed
preliminary simulations in order to choose the temperature and the
number of particles for the main simulations. In these initial
studies we found that finite size corrections are rather strong,
requiring a large number of particles to observe the correct
asymptotic form of the Casimir interaction at zero driving field. We
interpret this as a consequence of the unusual form of screening in
quasi-two-dimensional plasmas, coupled through the three dimensional
Coulomb interaction, which leads to a charge structure factor, in the
Debye-H\"uckel limit (for a single plate), of the form \cite{weis}
\begin{equation}
  S({\bf Q}) = \frac {|{\bf Q}|}{(|{\bf Q}| +m/2)}
\end{equation} with
$m = q^2 \bar n \beta/\epsilon$ {and $\bar{n} = N/H^2$. This singular form in $|{\bf Q}|$
  leads to a power-law decay of correlations in real space~\cite{weis},
  in contrast to the exponential decay of correlations in
  three-dimensional plasmas. We chose to run our simulations at a
  temperature such the Bjerrum length is comparable to the particle
  separation; specifically we took the values
  $T =1.03\, q^2 \sqrt{\bar n}/(4 \pi \epsilon )$. This corresponds to a
  region intermediate between the high and low temperature
  limits. Simulations at lower temperatures showed strong deviations
  from the Debye-H\"uckel structure factor, in particular $S({\bf Q})$ develops
  a weak  peak. At large temperatures the screening length $m^{-1}$  becomes large, physically  thermal fluctuations destroy screening, and the far field limit is pushed to distances $L$ comparable to the system size and thus into a region where finite size corrections become very important.

  We finally chose a series of systems with three different values
  of $N$: $N=1000, 2000, 4000$.  With these numbers of particles the
  un-driven systems shows a Casimir interaction, with the correct
  amplitude, over a range of separations from $0.02H$ to $0.12H$.  A
  simulation for $N=500$ showed strong modification of the Casimir
  amplitude for the separation of $0.02H$. Proper scaling of the data
  as the number of particles increases is a strong test of being in
  the continuum limit.

  For the values of $N$ that we study and separations below $0.02H$ we
  cross-over into a single particle regime, rather than a collective
  Casimir regime. For separations large than $0.12H$ there is an
  exponential decay of interactions, coming from the discrete nature
  of Fourier modes within a box. Within this range of distances
  Fig.~(2) in the main text shows that we control systematic errors at
  zero driving force to within about 2\%. The largest systems that we
  study require one month of simulation per point, to reduce
  statistical errors to within 2\%. The residual vertical spread in
  the points of Fig.~(2) is dominated by finite size corrections,
  which would require larger numbers of particles to reduce further.

  With a series of systems calibrated at zero driving force we
  performed the final series of simulations to measure interactions as
  a function of driving field. We note that at the largest driving
  field it is necessary to re-calibrate the criterion for the choice
  of $dt$: Relative motion of driven/un-driven particles is the
  fastest process in the simulation, which must be well resolved to
  generate accurate results.  The data presented is evaluated for an
  electric field which is skewed compared to the simulation
  cell. Other simulations (data not shown) show that the curves of
  Fig.~2 become anisotropic for the largest fields that we
  studied. This anisotropy can be understood by the fact that the
  driving reduces the effective number of interacting modes
  contributing to the Casimir interactions. This small number of modes
  in the summation then leads to angular modulation of the curves.
  This anisotropy can be treated analytically within the stochastic density functional theory by replacing the Fourier integrals in {Eqs.~(9, 10)}
  % (\ref{avf1},~\ref{avf2})
  by a discrete Fourier sum.
 
\subsection{Analytical results for the force}
The compact form of {Eqs.~(9, 10),} 
%(\ref{avf1},~\ref{avf2}), 
as derived from {Eq.~(8),} 
%(\ref{fq}), 
are deduced from the expressions
\begin{widetext}
  \begin{equation}
    \langle F_\perp(L)\rangle = -T S \int \!\! \frac{d^2{\bf Q}} {{(2\pi)}^2} \frac{Q \, m_1 m_2 \exp(-2QL)}{(m_1+2Q)(m_2+2Q)\left( 1+\frac{4 {({\bf Q}\cdot({\bf v}_1-{\bf v}_2))}^2}{Q^2{(D_1m_1+D_2m_2+2(D_1+D_2)Q)}^2} \right) - m_1 m_2 \exp(-2QL)}, 
\end{equation}
\begin{eqnarray}
 \langle {\bf F}_{||}(L)  \rangle &=& - T S \int  \frac{d^2{\bf Q}}{ {(2\pi)}^2}\frac{2 m_1 m_2 \, e^{-2 Q L} }{(m_1 + 2Q) (m_2 + 2Q) \left(1 +   \frac{4{({\bf Q}\cdot({\bf v}_1 - {\bf v}_2))}^2}{ Q^2   {(D_1 m_1 + D_2 m_2 + 2(D_1 + D_2) Q)}^2}\right) - m_1 m_2 \, e^{-2 Q L} }  \nonumber\\ && \times \frac{{({\bf Q}\!\cdot\!({\bf v}_1 - {\bf v}_2))} {\bf Q}}{Q(D_1 m_1 + D_2 m_2 + 2(D_1 + D_2) Q)},
\end{eqnarray}
\end{widetext}
which are derived from {Eqs.~(6) and (7)} 
%(\ref{frc1}) and (\ref{frc2}) 
under averaging and using
the density fluctuation correlation functions derived from {Eq.~(4).}
%(\ref{gbacefjk}).


\begin{thebibliography}{10}

\bibitem{bal07} P. Ball, Nature {\bf 447}, 772 (2007).

\bibitem{mos97} V. Mostepanenko, N. N. Trunov, {\sl The Casimir Effect
    and Its Applications}, (Oxford Science Publications), Oxford
  University Press, USA (1997).

\bibitem{Drosdoff} D. Drosdoff, I.V. Bondarev, A. Widom, R. Podgornik,
  L.M. Woods, Phys. Rev. X {\bf 6}, 011004 (2016).
  
 \bibitem{lif56} E. M. Lifshitz, Soviet Physics JETP {\bf 2}, 73
  (1956).

\bibitem{cas48} H. B. G. Casimir, Proc. Koninklijke Nederlandse
  Akad. Wetenschappen {\bf B51}, 793 (1948).

\bibitem{mil01} K A Milton, {\sl The Casimir Effect, Physical
    Manifestations of Zero-Point Energy}, World Scientific Publishing
  UK (2001).
\bibitem{dal11} D. Dalvit, P. Milonni, D. Roberts, F. Rosa, {\sl
    Casimir Physics}, Lecture Notes in Physics Volume 834,
  Springer-Verlag Berlin Heidelberg (2011).



\bibitem{par06} V. A. Parsegian, {\sl Van der Waals Forces}, Cambridge
  (Cambridge) (2006).


\bibitem{arn79}W. Arnold, S. Hunklinger, and K. Dransfeld,
  Phys. Rev. B, 19, 6049 (1979).

\bibitem{che07} F. Chen, G. L. Klimchitskaya, V. M. Mostepanenko, and
  U. Mohideen Phys. Rev. B {\bf 76}, 035338 (2007).

\bibitem{tor10}G. Torricelli, P. J. van Zwol, O. Shpak, C.  Binns,
  G. Palasantzas, B.J. Kooi, V. B. Svetovoy, M. Wuttig, Phys. Rev. A (R)
  {\bf 82} 010101 (2010).


\bibitem{tse12}W.-K. Tse and A. H. MacDonald, Phys. Rev. Lett. {\bf
    109}, 236806 (2012).

\bibitem{dor98}I.A. Dorofeyev, J. Phys. A: Math. Gen. {\bf 31}, 4369
  (1998).

\bibitem{ant05} M. Antezza, L.P. Pitaevskii and S. Stringari,
  Phys. Rev. Lett. {\bf 95},113202 (2005).

\bibitem{ant08} M. Antezza, L.P. Pitaevskii, S. Stringari and
  V.B. Svetovoy, Phys. Rev. A {\bf 77},{022901} (2008).
 
\bibitem{bim09}G. Bimonte, Phys. Rev. A. {\bf 80}, 042102 (2009).

\bibitem{kru11a} M. Kr\"uger, T. Emig, G. Bimonte and M. Kardar,
  Europhys. Lett. {\bf 95},21002 (2011).

\bibitem{kru11b} M. Kr\"uger, T.  Emig and M. Kardar,
  Phys. Rev. Lett.{\bf 106}, 210404 (2011).

\bibitem{mes11} R. Messina and M. Antezza, Europhys. Lett. {\bf 95},
  61002 (2011).

\bibitem{rod11} A.W. Rodriguez, O. Ilic, P. Bermel, I. Celanovic,
  J.D. Joannopoulos, M. Solja\c{c}i\'{c} and S.G. Johnson,
  Phys. Rev. Lett. {\bf 107} 114302 (2011).

\bibitem{lu15} B.-S. Lu, D.S. Dean, and R. Podgornik,
  Europhys. Lett. 112, 20001 (2015).

\bibitem{jaf05} R.L. Jaffe, Phys. Rev. D {\bf 72}, 021301(R) (2005).

\bibitem{cug12} J. Cugnon, Few-Body Syst. {\bf 53}, 181 (2012).

\bibitem{lev99}Y. Levin, Physica A {\bf 265}, 432 (1999).

\bibitem{bue05} P.R. Buenzli and Ph. A. Martin, Europhys. Lett. {\bf
    72} 42 (2005).
 
\bibitem{jan05} B. Jancovici and L. Samaj, Europhys. Lett. {\bf
    72}{35} (2005).
    
\bibitem{comm1} If the  circuit of  plate 2 is closed, a current is induced in the plate 2  by 
 the {\em Coulomb drag} due to plate 1, this is the case considered here. If the circuit of the opposing plate is open, charge accumulates at the plate edges, creating an in-plane electrical field which cancels the due to the driven plate and prevents current flow in the opposing plate.  This open configuration can be studied both numerically and analytically and is left for further study.

\bibitem{comm2} An electric current in one
  plate will lead to a steady state where this plate will have a
  higher temperature than the other. A difference in temperature can be taken into
  account in both our numerical and analytic treatment~\cite{lu15} of
  the problem, however for the sake of simplicity it is not studied in
  this letter.

\bibitem{yeh}I.-C. Yeh and M. L. Berkowitz, J. Chem. Phys.  {\bf 111},
  3155 (1999).

\bibitem{SM} For further details see supplementary material, which includes Refs. \cite{errors,reverse,quantum,weis}.

\bibitem{dea96}D.S.  Dean, J. Phys. A: Math. Gen. {\bf 29}, 613
0  (1996).

\bibitem{kaw94}K. Kawasaki, Physica A {\bf 208}, 35 (1994).

\bibitem{dea14}D.S. Dean and R. Podgornik, Phys. Rev. E 89, 032117
  (2014).

\bibitem{zwa01} R. Zwanzig, {\sl Non equilibrium statistical
    mechanics}, Oxford University Press, Oxford (2001).

\bibitem{dem15} V. D\'emery and D.S. Dean, {J. Stat. Mech. {\bf 2016}, 023106 (2016).}

\bibitem{dem10}V. D\'emery and D.S. Dean, Phys. Rev. Lett. {\bf 104}, 080601 (2010). 
\bibitem{dem11} V. D\'emery and D.S. Dean, Phys. Rev. E {\bf 84}, 010103(R) (2011)

\bibitem{hem96}P.C. Hemmer, H. Holden and S. Kjelstrup Ratkje
  eds. {\sl The collected works of Lars Onsager (with commentary)},
  World Scientific (Singapore) (1996).
\bibitem{errors}
H. Flyvbjerg and H. G. Petersen, J. Chem. Phys. {\bf 91}, 461(1989).

\bibitem{reverse}
A. S. Kronfeld, Progress of Theoretical Physics Supplement,   {\bf 111}, 293 (1993).


\bibitem{quantum} E. Wigner Phys. Rev. {\bf 40}, 749, (1932).


\bibitem{weis}J.J. Weis, D. Levesque, and J.M. Caillol,
  J. Chem. Phys., {\bf 109}, 7486, (1998).
\end{thebibliography}

\begin{thebibliography}{10}


\bibitem{errors}
H. Flyvbjerg and H. G. Petersen, J. Chem. Phys. {\bf 91}, 461(1989)

\bibitem{reverse}
A. S. Kronfeld
%http://cds.cern.ch/record/546976/files/9205008.pdf %error analysis
 Progress of Theoretical Physics Supplement  (1993) {\bf 111}, 293-311.


\bibitem{quantum}
%http://journals.aps.org/pr/pdf/10.1103/PhysRev.40.749
%On the Quantum Correction For Thermodynamic Equilibrium
E. Wigner Phys. Rev. {\bf 40}, 749, (1932).


\bibitem{weis}J.J. Weis, D. Levesque, and J.M. Caillol,
  J. Chem. Phys., {\bf 109}, 7486-7497, (1998).
  % doi = http://dx.doi.org/10.1063/1.477371



\end{thebibliography}
\end{document}